\documentclass[apj]{emulateapj}
\usepackage{apjfonts}

\slugcomment{Accepted to {\it{The Astrophysical Journal Letters}}}
\shortauthors{Gonzalez et al.}
\shorttitle{Galaxy Cluster Assembly at $z=0.37$}

\begin{document}

\newcommand{\kms}{~km~s$^{-1}$~}
\newcommand{\logh}{$+5\log h_{70}$}
\newcommand{\chandra}{{\it{Chandra~}}}
\newcommand{\ho}{~}
\newcommand{\msun}{M$_\odot$~}
\newcommand{\sg}{SG1120-1202}
\title{Galaxy Cluster Assembly at $z=0.37$$^{1}$}

\author{Anthony H. Gonzalez$^{2,3}$, Kim-Vy H. Tran$^4$, Michelle N. Conbere$^3$, 
\& Dennis Zaritsky$^5$ }

\footnotetext[1]{Based on observations with the Chandra X-ray
  Observatory, the VLT (program 072.A-0367), the Magellan Baade telescope, 
and the MMT Observatory, a joint venture of the Smithsonian Astrophysical 
Observatory and the University of Arizona.}
\footnotetext[2]{NSF Astronomy and Astrophysics Postdoctoral Fellow}
\footnotetext[3]{Department of Astronomy, University of Florida,
  Gainesville, FL 32611}
\footnotetext[4]{Institute for Astronomy, ETH Z\"urich, CH-8093
  Z\"urich, Switzerland}
\footnotetext[5]{Steward Observatory, University of Arizona, 933 
North Cherry Avenue, Tucson, AZ 85721}

\begin{abstract}

  We present X-ray and spectroscopic confirmation of a cluster assembling from
  multiple, distinct galaxy groups at $z=0.371$.  Initially detected in the
  Las Campanas Distant Cluster Survey, the structure contains at least four
  X-ray detected groups that lie within a maximum projected separation of 4
  Mpc and within $\Delta \mathrm{v}=550$\kms of one another.  Using \chandra
  imaging and wide-field optical spectroscopy, we show that the individual
  groups lie on the local $\sigma-T$ relation, and derive a total mass of
  $M\ge 5\times 10^{14}$ \msun for the entire structure.  We demonstrate that
  the groups are gravitationally bound to one another and will merge into a
  single cluster with $\gtrsim\frac{1}{3}$ the mass of Coma.  We also find
  that although the cluster is in the process of forming, the individual
  groups already have a higher fraction of passive members than the field.
  This result indicates that galaxy evolution on group scales is key to
  developing the early-type galaxies that dominate the cluster population by
  $z\sim0$.
  
\end{abstract}


\section{Introduction}

A fundamental prediction of hierarchical structure formation is that galaxy
clusters assemble at late times from the merging and accreting of smaller
structures \citep{peebles:70}.  While many examples of the accretion of groups
and galaxies onto massive clusters exist
\citep[e.g.][]{abraham:96,hughes:98,tran:05}, we lack clear early-stage
examples of clusters being assembled from an ensemble of galaxy groups.  The
identification of such protoclusters is challenging due to the extensive
mapping required to trace the substructure that will eventually form the
cluster.

Nonetheless, detection of these rare systems is well worth the effort.  They
provide a detailed snapshot of the process of filamentary collapse and
assembly in the quasi-linear regime, and enable a {\it direct} study of a
class of progenitors that does not yet include a massive cluster.  The latter
is particularly critical for determining if ``pre-processing'' in groups is
responsible for the bulk of the observed differences in morphology and stellar
populations between cluster and field galaxies \citep{zabludoff:98,kodama:01}

We present confirmation of a protocluster identified optically in the Las
Campanas Distant Cluster Survey \citep[LCDCS;][]{gonzalez:01}.  Designed to
provide a catalog of clusters at $z\approx0.35-0.9$, the LCDCS extends down to
group masses at the lowest redshifts and hence can be used to identify
filamentary structures at $z\sim0.4$.  One such candidate structure includes
four LCDCS candidates within a 3.5$\arcmin$ radius region with estimated
redshifts of $z_{est}=0.35-0.50$.  We demonstrate that we are witnessing the
assembly of a cluster from what can be described in analogy to superclusters
as a supergroup. Throughout this paper, we use a standard cosmology
($\Omega_M=0.3$, $\Omega_\Lambda=0.7$, $H_0=70$\kms).

\vskip 0.5cm
\section{X-ray Observations with Chandra}

The target field was observed with \chandra\ ACIS-I in Very Faint (VF) mode on
March 24, 2002 (Obs ID 3235) for 70.13 ks, with the detector oriented such
that one LCDCS candidate was imaged in each ACIS-I chip.  Data reduction was
performed in the standard fashion using CIAO 3.0.  The level 1 event file was
reprocessed to correct for charge transfer inefficiency and to reduce the VF
mode particle background,\footnote{See
\url{http://cxc.harvard.edu/cal/Acis/Cal\_prods/vfbkgrnd/} } and bad pixels
and events with bad grades were removed in the standard fashion.  Strong
flares were removed using the $\mathrm{lc\_clean}$ routine written by Maxim
Markevitch, yielding a net exposure time of 64 ks. The high energy background
in the remaining time interval is consistent with the quiescent rate from the
"blank-sky" background file to within 1\%.\footnote{See
\url{http://cxc.harvard.edu/ciao/threads/acisbackground/}}

\begin{figure}
\epsscale{1.0}
\plotone{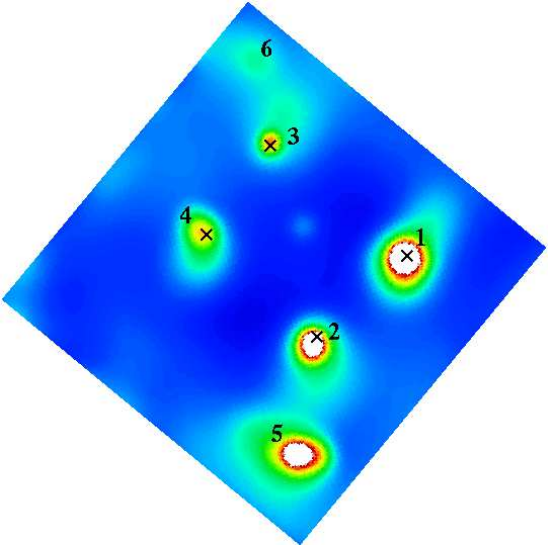}
\caption{Smoothed X-ray map ($13.2'\times 13.2'$) of the structure with
  point sources excised; north is up and east is to the left. The
  crosses denote the locations from the LCDCS catalogs and the numeric
  labels correspond to the designations in Table \ref{tab:tab1}.
  Redshifts have been obtained for sources 1-5.
\label{fig:xray}}
\end{figure}

Image analysis is performed in the 0.8-3 keV band. Using wavelet detection, we
identify six distinct extended sources at $>5\sigma$ significance (Figure
\ref{fig:xray}). Four of these sources are coincident to within $10\arcsec$
with LCDCS candidates, while the other two extended sources have no LCDCS
counterparts. One of these latter two (source 5) could not have been detected
by the LCDCS due to its proximity to a bright star; the other is the faintest
of the six sources and appears to be a very poor group.

Spectral analysis is performed in the 0.7-8 keV band.  Temperatures are
measured within $400$\ho kpc circular apertures to maximize signal-to-noise
(S/N), excluding point sources, and using the appropriately normalized ACIS
"blank-sky" file for the background.  We fit the spectra using a MEKAL model,
with the absorption fixed to the galactic value $n_H=5.13\times10^{20}$
cm$^{-2}$ and the metal abundance fixed at $Z=0.3$ \citep{mushotzky:1997}.
For four of the extended sources, we fix the redshift to the median velocity
of the associated galaxies (see \S \ref{sec:spec}), while for LCDCS 0259
(source 2), which lies outside the spectroscopic survey region, we use the
single redshift obtained for the brightest associated galaxy.  The best-fit
temperatures (Table \ref{tab:tab1}, with 68\% uncertainties) range from
$T=1.7-3$ keV and are consistent with the temperatures found for groups or
poor clusters similar to Virgo \citep{shibata:01}.  Only for the faintest
source do we lack sufficient S/N to obtain a temperature.

\section{Optical Spectroscopy}
\label{sec:spec}

\subsection{MMT, Magellan, and VLT Observations}

To confirm that the X-ray sources are part of a larger structure, long-slit
spectra of the candidate brightest group galaxies (BGGs) associated with each
of the six X-ray peaks were obtained with Magellan and the MMT in February and
April 2003, respectively.  Redshifts obtained for the BGG's in groups 1-5
confirm that at least four of the X-ray regions are associated with a single
physical complex at $z=0.37$.  Only the BGG candidate for group 1 was found to
lie at a different redshift ($z=0.48$).  These initial results motivated a
more extensive spectroscopic program with the VLT.

\begin{figure}
\epsscale{1.0}
\plotone{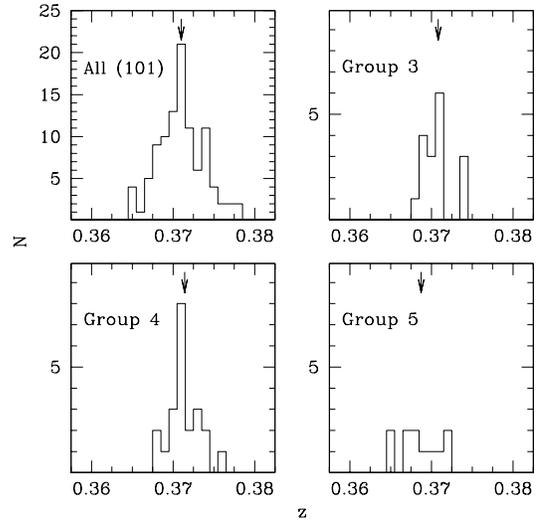}
\caption{{\it Top left} --
  Redshift distribution of the 101 confirmed members (defined by $0.36<z<0.38$)
  in \sg~($z=0.3710\pm0.0004$).  Although the members trace a very
  large structure (projected spatial distance $\sim4$\ho Mpc) and
  include galaxies from 3 of the groups at $z=0.37$, they have a
  remarkably narrow redshift distribution and a 
  velocity dispersion of only $\sigma=616\pm50$\kms. 
  {\it Top right and lower panels} -- Redshift distributions for the individual groups; the
arrow denotes the average redshift of the members.  To isolate
  members of each individual group, here we consider only members
  within 500\ho kpc of the X-ray centers.  Even with our
  conservative selection criterion, the maximum redshift difference
  between the groups is only $\Delta z=0.0025$. 
\label{fig:zhist1}}
\end{figure}

Using VIMOS \citep{lefevre:03}, we measure redshifts and determine membership
for galaxies near groups 1, 3, 4, and 5; group 2 was excluded due to an
interchip gap.  From a $14'\times16'$ $R$-band image (1820s) taken with VIMOS,
we generated a catalog of $\sim2100$ objects with $R\leq22.5$.  Our
observations used three multi-slit masks with $1''-$wide slits and targeted
443 of these objects; each mask had a total integration time of 2400 seconds.
While ours is a magnitude limited survey and targets were not selected by
morphology, preference was given to objects in visually overdense regions.
The intermediate resolution (MR) grism on VIMOS gave us a spectral range of
$0.5-1.0\mu$m and spectral resolution of 12.2\AA.

To reduce and extract spectra, we use a combination of IRAF\footnote{IRAF is
distributed by the National Optical Astronomy Observatories, which are
operated by the Association of Universities for Research in Astronomy, Inc.,
under cooperative agreement with the National Science Foundation.} routines
and custom software provided by D. Kelson \citep{kelson:98}; a more detailed
explanation of the reduction pipeline can be found in \citet{tran:05}.  We use
a spectrophotometric standard to correct for the telluric A and B bands and
flux calibrate the spectra.

We determine redshifts with the IRAF cross-correlation routine XCSAO
\citep{kurtz:92}.  Our final redshift catalog has 413 objects, corresponding
to a target success rate of 93\%.  The average redshift uncertainty estimated
by XCSAO is $\sim30$\kms; while XCSAO tends to underestimate the errors, this
problem does not affect our results.

\subsection{Redshift Distribution}
\label{sec:redshifts}

From spectroscopy of 4 of the 6 regions with extended X-ray emission, we
identify a peak corresponding to LCDCS 0258 ($z=0.48$, source 1), and another
corresponding to three X-ray groups at $z=0.37$, hereafter referred to as
\sg~(SG standing for ``supergroup''). The mean redshift of the 101 galaxies
belonging to \sg~ is $z=0.3710\pm0.0004$.  Although the members lie in
multiple, distinct X-ray bright regions and trace a large structure with a
projected spatial distance of $\sim4$\ho Mpc, their redshift distribution is
remarkably narrow (Fig.~\ref{fig:zhist1} and Table~\ref{tab:tab1}) and their
velocity dispersion surprisingly small ($\sigma=616\pm50$\kms).  The mean
redshift, velocity dispersion, and associated errors are determined using the
biweight and jacknife methods \citep{beers:90}.  To compare the dynamics of
the three X-ray groups to each other, members of each group are selected as
galaxies that lie within 500\ho kpc of their respective X-ray peak.  The
redshift distributions of the three groups are shown in Figure
\ref{fig:zhist1} and their velocity dispersions are listed in Table
\ref{tab:tab1}.

\subsection{Spectral Populations in \sg}

To quantify the recent star formation histories of galaxies in \sg, we
separate the members by [OII] equivalent width into absorption
([OII]~$\lambda3727<5$\AA; "passive") and emission
([OII]~$\lambda3727\geq5$\AA; "active") line galaxies.  We use the same
bandpasses as in \citet{fisher:98} to measure the [OII] doublet, and our
wavelength coverage includes the [OII] doublet for 95 of the 101 confirmed
members.

The fraction of passive galaxies in \sg~ is $61\pm8$\%.  This fraction is less
than in CL~1358+62 ($81\pm6$\%), a more massive cluster at comparable redshift
\citep[$z=0.33$, $\sigma=1027$\kms;][]{fisher:98}, but twice as high as the
field value \citep[$27\pm4$\% at $0.2<z<0.5$;][]{tran:04b}. The passive galaxy
fraction in \sg~ most closely resembles that found in X-ray luminous groups in
the nearby universe \citep[$69\pm7$\%;][]{tran:01}.  Direct comparison to
these other samples is valid because all four are magnitude-selected and use
the same [OII] selection criteria.

\begin{deluxetable}{llllllll}
\tabletypesize{\scriptsize}
\tighten
\tablecaption{Group Sample}
\tablewidth{0pt}
\tablehead{
\colhead{ID} &
\colhead{$\alpha$ } &
\colhead{$\delta$ } &
\colhead{LCDCS} &
\colhead{$z$} &
\colhead{$T$} &
\colhead{$\sigma$ } &
\colhead{$N_{g}$\tablenotemark{a}}\\
\colhead{} &
\colhead{(J2000)} &
\colhead{(J2000)} &
\colhead{ID} &
\colhead{} &
\colhead{(keV)} &
\colhead{\kms} &
\colhead {} 
}
\startdata
1 & 11:19:55.2  &$-$12:02:28 & 0258 & 0.4794 &    $2.3^{+0.4}_{-0.3}$ & $820\pm101$ & 17  \\
2 & 11:20:07.6  &$-$12:05:13 & 0259 & 0.3707&  $2.2^{+0.7}_{-0.4}$ & --- & 1   \\
3 & 11:20:13.2  &$-$11:58:44 & 0260 & 0.3704 &    $1.7^{+0.5}_{-0.3}$ & $369\pm76$ & 17 \\
4 & 11:20:22.3  &$-$12:01:45 & 0264 & 0.3713 &    $1.8^{+1.2}_{-0.5}$ & $446\pm83$ & 22 \\
5& 11:20:10.0  &$-$12:08:50 & ---  & 0.3688 &    $3.0^{+1.2}_{-1.0}$ & $557\pm93$ & 11  \\
6 & 11:20:15.6  &$-$11:56:03 & --- &  ---  &    ---                & ---  & --- \\
\enddata
\tablenotetext{a}{Number of galaxies used to calculate the mean redshift and velocity dispersion.}
\label{tab:tab1}
\end{deluxetable}

\section{Cluster Assembly}

\sg's extended, roughly linear geometry is seen in both its X-ray emission and
the spatial distribution of its galaxies.  Coupled with the small dispersion
in the mean group redshifts ($\Delta z=0.0025$) and visible signs of
interaction in the X-ray map (groups 2+5 and 3+6), this geometry argues that
we are witnessing the initial stages of filamentary collapse and the formation
of a more massive cluster.  To test this hypothesis, we estimate \sg's total
mass, assess the probability that the system is bound, and estimate the
dynamical timescale for the system to merge.  We include groups 2-5 in this
analysis; we are confident that group 2 is part of the complex despite having
only one redshift, both because the redshift is for the BGG (a 3 $L_*$
elliptical within 3$\arcsec$ of the X-ray peak) and because asymmetry in the
X-ray map indicates that groups 2 and 5 are interacting.

\subsection{Total Mass}

\begin{figure}
\epsscale{1.0}
\plotone{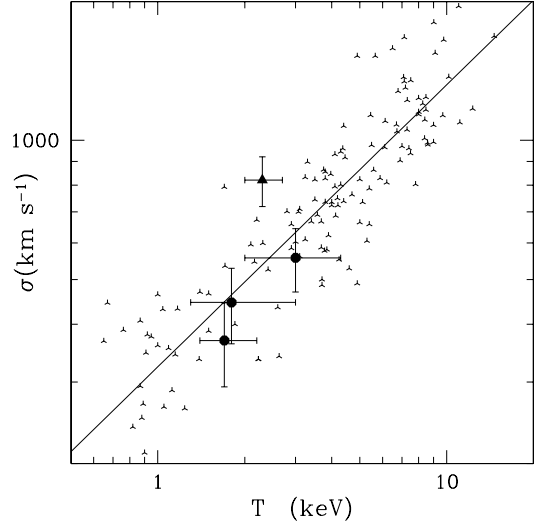}
\caption{ Comparison of the groups in this study with the local $\sigma-T$
  relation \citep{xue:00}. The solid symbols with error bars correspond to the
  groups in our sample, circles denoting members of the supergroup ($z=0.37$)
  and the triangle LCDCS 0258 ($z=0.48$, source 1). The other points are the local data
  from \citet{xue:00} and the solid line is their best fit for this combined
  group and cluster sample. Our sample is consistent with this local relation.
\label{fig:sigt}}
\end{figure}

Given the consistency of the groups with the local $\sigma-T$ relation (Figure
\ref{fig:sigt}), we assume hydrostatic equilibrium and use the local
mass-temperature ($M-T$) relation in \citet[][Equation 16]{shimizu:03} to
estimate the masses of the individual groups.\footnote{Evolution in the $M-T$
relation should be negligible over this redshift interval \citep{ettori:04}.}
The four spectroscopically confirmed groups at $z=0.37$ have $T=1.7-3$ keV,
yielding a combined mass of $M=5.3^{+1.6}_{-0.9}\times10^{14}$ \msun.  This
mass is a lower limit on the total mass of the system because it does not
include group 6, groups that may lie outside the \chandra~field, or material
in filaments between groups.  It is equivalent to that of a single cluster
with $T\sim5$ keV, and at least a third the mass of the Coma cluster
\citep{hughes:1989,honda:1996}.

\subsection{Dynamical State}

That the individual groups in \sg~lie on the local $\sigma-T$ relation argues
that each group is approximately virialized and that, if the larger structure
is bound, the groups are infalling for the first time.  We perform a simple
dynamical analysis to assess the likelihood that the \sg~complex is indeed
gravitationally bound.

The individual groups will be bound to the larger system if 
\begin{equation}
\frac{GM}{R_{p} (\sin i)^{-1}} \geq \frac{\mathrm{v}_{pec}^2}{2}
\end{equation}
where $M$ is the total mass, $\mathrm{v}_{pec}$ is the pairwise peculiar
velocity for groups, $R_{p}$ is the projected distance between the groups, and
$i$ is the opening angle.  We assume the local value of
$<\mathrm{v}_{pec}>=325\pm175$ \kms from \citet{padilla:01} and conservatively
take $R_{p}= 3$\ho Mpc; this value corresponds to the largest separation
between the X-ray groups.

Using our X-ray estimate of the total mass, the above inequality holds for
$i\ga4^\circ$. Geometrically, the probability is 99.5\% that the opening angle
is at least this large. Even assuming $\mathrm{v}_{pec}=500$\kms ($1\sigma$
above the mean value from \citealt{padilla:01}), the inequality holds for
$i\ga10^\circ$ and the probability that the groups are bound remains high
(97\%).

The above analysis indicates that an unbound system is unlikely, but does not
conclusively eliminate the possibility.  We can strengthen the constraint by
exploring the implications of an unbound system.  For $i \la 10^\circ$, the
groups are separated by $\ga20$ Mpc, and the corresponding Hubble flow
redshift offset is $(\Delta z)_H\ga0.0075$ (2250 \kms). In this case, $(\Delta
z)_H$ is a factor of three greater than the observed maximum redshift
separation between the groups.  Thus relative peculiar velocities of $\ga1500$
\kms would be required to counterbalance the Hubble flow and explain the small
observed redshift separation; this is an implausibly large value if the groups
are truly separated by $\ga20$ Mpc and not infalling.

We conclude that the groups in \sg~are bound to each other.  The dynamical
time for this system is
\begin{equation}
t_{dyn}\approx \sqrt{\frac{R^3}{GM}} \approx 1.2 (\sin i)^{-3/2} \;\;\mathrm{Gyrs},
\end{equation}
for $R_{p}=1.5$\ho Mpc and $M=5\times 10^{14}$ \msun.  For $i\ga26^{\circ}$,
$t_{dyn}$ is less than the lookback time.

An alternative treatment is to assume that the groups are bound and on radial
orbits that have not yet crossed the center of the potential, and use the
timing argument \citep{kahn:1959} to determine their line-of-sight distance by
fitting the projected separation and radial velocity in the center of mass
frame. For the four groups we find solutions with line-of-sight distances from
the center-of-mass of between 7.5 and 10 Mpc.  These solutions have $11^\circ
< i < 27^\circ$, and hence suggest that the complete collapse of the system
will occur after the current time.  The main conclusion from this analysis is
that a bound solution exists with the measured masses that can reproduce the
positions and radial velocities of the groups. Detailed descriptions of the
orbits and future merger history require more information.

\section{Conclusions}

Our observations confirm that \sg~is a protocluster being built up by the
assembly of multiple galaxy groups.  From deep \chandra imaging and optical
spectroscopy, we find that \sg~ contains a minimum of four X-ray luminous
groups at $z=0.371$ within a projected 4\ho Mpc of one another.  The groups
have X-ray temperatures of $T=1.7-3$ keV and their mean redshifts span a
narrow range of $\Delta z=0.0025$ (550\kms).  The group velocity dispersions
and X-ray temperatures fall on the local $\sigma-T$ relation, indicating that
they are each virialized systems with a combined mass $M\ge5\times 10^{14}$
\msun.  Using dynamical arguments, we demonstrate that the X-ray groups are
gravitationally bound to one another and should merge into a single cluster;
the resulting cluster will have $\gtrsim\frac{1}{3}$ the mass of Coma.

While \sg~represents only one of a variety of accretion histories that can
yield a moderate mass cluster at $z=0$, this assembly path -- the late merging
of multiple roughly equal mass subhalos -- is of particular interest.
\sg~provides a unique laboratory for assessing the impact of local density
upon the evolution of member galaxies, and for quantifying the degree to which
the group environment acts to ``pre-process'' galaxies prior to cluster
assembly.  Already we see that the individual groups have twice as many
passive galaxies as the field, indicating that galaxy evolution on group
scales is key to developing the early-type galaxies that dominate the cluster
population by $z\sim0$.

\acknowledgements

We are indebted to A. Vikhlinin and M. Markevitch for their assistance
and for providing us with codes employed in the \chandra analysis.  Support
for this work was provided by NASA through Chandra Award GO2-3183X3 issued
by the CXO Center, which is operated by SAO for and on behalf of NASA under
contract NAS8-03060. AHG is funded by an NSF Astronomy \& Astrophysics
Postdoctoral Fellowship under award AST=0407085, KT acknowledges support from
the Swiss National Science Foundation, and DZ acknowledges fellowships from
the David and Lucile Packard Foundation and Alfred P. Sloan Foundation.

%Facilities: 
%   \facility{CXO(ACIS)}, \facility{VLT(VIMOS)}, 
%   \facility{MMT}, \facility{Magellan}

\bibliographystyle{apj}
\bibliography{ms}

\end{document}